\documentclass[usenatbib,usegraphicx,useAMS]{mn2e}
\usepackage{amsmath}
\usepackage{amssymb}

\def\cf{{cf.}} 
\def\eg{{e.g.}} 
\def\ie{{i.e.}} 

\def\s{{\rm s}} 
\def\ms{{\rm m}\s} 
\def\Ms{{\rm M}\s} 

\def\m{{\rm m}} 
\def\cm{{\rm c}\m} 
\def\km{{\rm k}\m} 

\def\g{{\rm g}} 
\def\Ms{M_\odot} 

\def\eV{{\rm eV}} 
\def\keV{{\rm k}\eV} 
\def\GeV{{\rm G}\eV} 
\def\erg{{\rm erg}} 

\def\G{{\rm G}} 

\renewcommand{\d}{{\rm d}}
\newcommand{\e}{{\rm e}}

\def\RNS{R_{\rm NS}}
\def\MNS{M_{\rm NS}}
\def\BNS{B_{\rm NS}}
\def\MBH{M_{\rm BH}}
\def\BBH{B_{\rm BH}}
\def\dsn{\rho_{\rm sn}}
\def\vsn{v_{\rm sn}}
\def\MHe{M_{\rm He}}
\def\RHe{R_{\rm He}}

\title{Supernovae in Helium Star--Compact Object Binaries: A Possible
$\gamma$-ray Burst Mechanism}
\author[Avery E. Broderick]{Avery E. Broderick\thanks{E-mail:
abroderick@cfa.harvard.edu}\\
Institute for Theory and Computation, Harvard-Smithsonian Center for Astrophysics, 60 Garden St., MS 51, Cambridge, MA 02138, USA\\}

\begin{document}
\maketitle

\begin{abstract}
Helium star--compact object binaries, and helium star--neutron star
binaries in particular, are widely believed to be the
progenitors of the observed double neutron star systems.  In these, the
second neutron star is presumed to be the compact remnant of the
helium star supernova.  Here, the observational implications of such a
supernova are discussed, and in particular are explored as a candidate
$\gamma$-ray burst mechanism.  In this scenario the supernova
results in a transient period of rapid accretion onto the compact object,
extracting via magnetic torques its rotational energy at highly
super-Eddington luminosities in the form of a narrowly beamed,
strongly electromagnetically dominated jet.  Compton scattering of
supernova photons advected within the ejecta, and photons originating
at shocks driven into the ejecta by the jet, will cool the jet and can
produce the observed prompt emission characteristics, including the
peak--inferred isotropic energy relation, X-ray flash characteristics,
subpulse light curves, energy dependent time lags and subpulse
broadening, and late time spectral softening.  The duration of the
burst is limited by the rate of Compton cooling of the jet, eventually
creating an optically thick, moderately relativistically expanding
fireball which can produce the afterglow emission.  If the black hole
or neutron star stays bound to a compact remnant, late term light
curve variability may be observed as in \mbox{SN 2003dh}.
\end{abstract}


\section{Introduction} \label{I}
Despite being discovered more than 30 years ago
\citep{Kleb-Stro-Olso:73}, and the fact that bursts have been detected
daily since the advent of the Compton Gamma-Ray Observatory in 1991,
there still does not exist a satisfactory theoretical model which
encompasses the entire $\gamma$-ray burst event.  Nonetheless, in
spite of their strong heterogeneity, the abundance of observations has
led to a well described burst phenomenology, and is briefly summarised
below.

Burst durations are bimodal, with long bursts lasting $\sim100\,\s$
while short bursts last $\sim1\,\s$ \citep{Kouv-etal:93}.  Long bursts typically
have inferred isotropic luminosities on the order of $\sim
10^{51-52}\,\erg/\s$ \citep[see, \eg,][and references
  therein]{Piran:04}, although achromatic breaks in the afterglow
light curves suggest strong collimation \citep[with opening angles
  $\sim5-10^\circ$,][]{Frai-etal:01} which, in turn, implies actual
luminosities on the order of $\sim 10^{48-49}\,\erg/\s$.  The majority
of the prompt emission is in the form of apparently nonthermal
$\gamma$-rays, well fit by a broken power law, typically peaking near
$100$--$1000\,\keV$, and softening throughout the burst. The prompt
emission is composed of a large number of subpulses with typical
widths on the order of a second in which lower energy emission lag
behind, and are wider than, the higher energy emission.

Long bursts are followed by optical and radio afterglows thought to be
generated by a hot fireball interacting with the interstellar medium
\citep[see, \eg,][]{Mesz:02,Li-Chev:03}.  These can
last many weeks and have in at least two cases (\mbox{GRB 980425} and
\mbox{GRB 030329}) included type Ib,c supernova (\mbox{SN 1998bw} and
\mbox{SN 2003dh}, respectively) light curves $7$--$10\,{\rm days}$
after the prompt emission.  The total bolometric energy in these
bursts has been estimated from late time radio observations to be
$\sim 10^{51}\,\erg$ and roughly constant among bursts
\citep{Berg-Kulk-Frai:04}.  To date, there have been no instances of
afterglow or supernova being associated with short bursts.

Due to similarities in their temporal structure, duration, and
spectra, X-ray flashes appear to be low luminosity cousins of
$\gamma$-ray bursts.  Recently it has been shown that their bolometric
luminosity is indeed comparable to normal bursts \citep{Sode-etal:04}.
Despite this, their inferred isotropic energy is substantially less
($\sim10^{49}\,\erg$) placing X-ray flashes upon the peak--inferred
isotropic energy relation discovered by \citet{Amat-etal:02},
and supporting a unified interpretation of X-ray flashes and
$\gamma$-ray bursts \citep{Saka-etal:04}.

A number of mechanisms have been suggested for $\gamma$-ray bursts.
Frequently, these address either the central engine and emission
separately.  Central engine models include black hole and/or neutron star
collisions
\citep{Eich-Livi-Pira-Schr:89,Pacz:91,Nara-Pacz-Pira:92},
magnetar birth \citep{Usov:92,Thom:94,Thom-Chan-Quat:04}, black hole
birth \citep{Viet-Stel:98}, and collapsar
\citep[see, \eg,][]{MacF-Woos-Hege:01} models.  In most of these it is
not clear how the spectral and temporal structure of the burst is
produced.  In contrast, there are also a number of models which focus
upon the emission.  These include the cannonball
\citep[][and references therein]{Dado-Dar-DeRu:02},
shot gun \citep{Hein-Bege:99}, Comptonised jet
\citep{Thom:94,Ghis-Lazz-Celo-Rees:00}, and internal shock
\citep{Piran:04} models.  Despite their ability to reproduce many of
the observed spectral and temporal features of the bursts, few of
these address directly the power source of the burst.  As a result,
currently there are few models which can explain both the properties
of the observed emission and the prodigious energy of observed bursts
in a satisfactory manner.

\begin{figure*}
\begin{center}
\includegraphics[width=1.5\columnwidth]{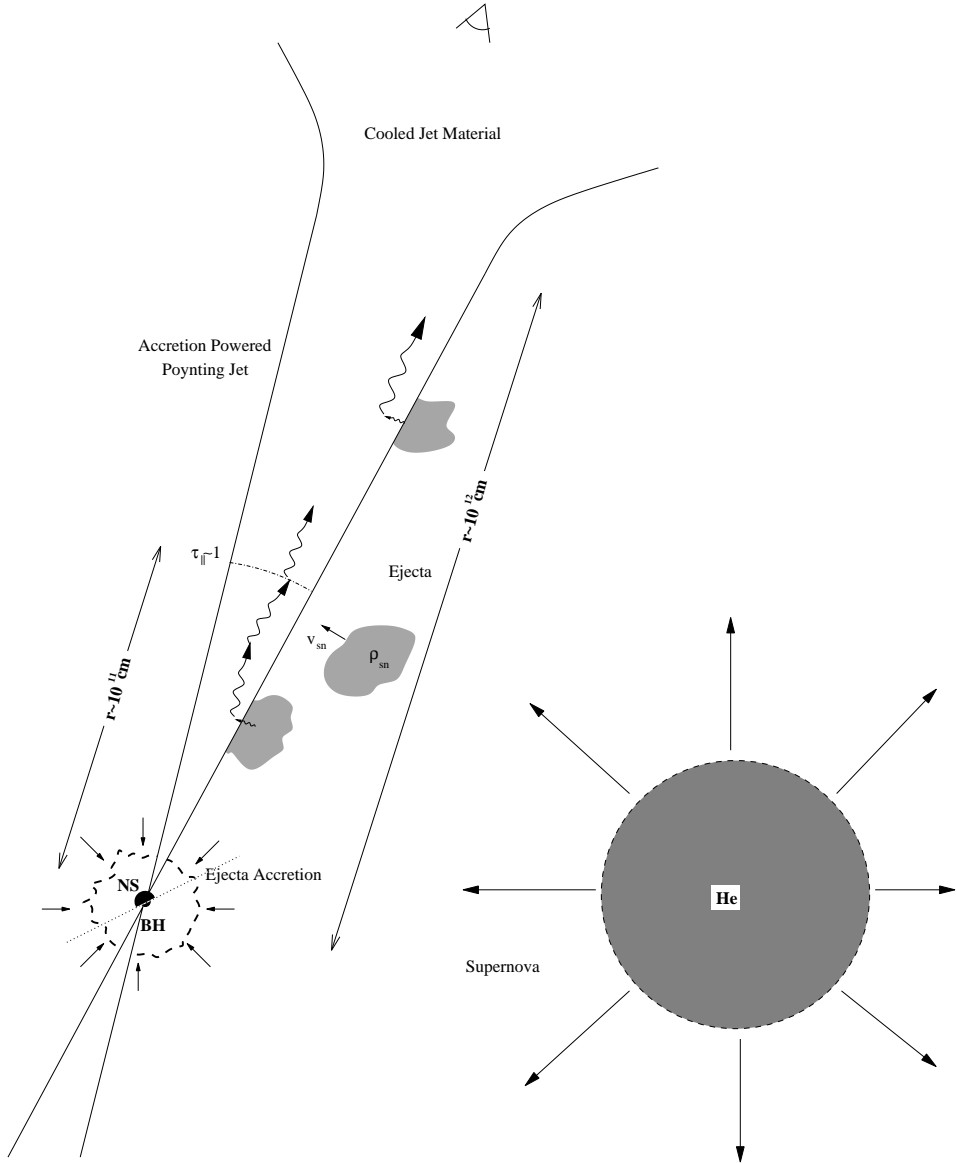}
\caption{A cartoon of the mechanism is shown (not to scale) with the
  principle features, supernova ejecta, accretion driven jets,
  up-scattered entrained and shock produced photons from the ejecta,
  and the regions where the jet is axially optically thin
  ($r\sim10^{11}\,\cm$) and becoming nonrelativistic.}
\label{cartoon}
\end{center}
\end{figure*}

Here a model involving supernovae in helium star--black hole and
helium star--neutron star binaries is presented.  In this scenario the
supernova results in a transient period of rapid accretion onto the compact object, extracting
via magnetic torques its rotational energy at highly super-Eddington
luminosities in the form of a narrowly beamed, strongly
electromagnetically dominated jet.  The prompt emission is produced by
Compton scattering supernova photons advected within the ejecta and
photons created at shocks driven into the ejecta by the jet.  The
duration of the burst is limited by the rate of Compton cooling of the
jet, eventually creating an optically thick, moderately
relativistically expanding fireball which can produce the afterglow
emission.  In comparison with the previous models this has two
advantages: firstly it provides a unified model for the central engine
and the prompt emission, and secondly, since helium star--neutron star
binaries are widely believed to be the progenitors of the observed
double pulsars, it is assured that supernova in these systems will
occur in sufficient quantity (see section \ref{PE:PS}).  For
convenience, typical values for the pertinent quantities which
describe the model are collected in Table \ref{quantities} and a
schematic of the process is shown in Figure \ref{cartoon}.

The formation and energetics of the jet are discussed in Section \ref{JF}.  Sections \ref{PE} and \ref{Ag}
describe the mechanisms by which the kinetic energy of the jet is
converted in the the observed prompt emission and subsequent
afterglow, including the limits this places upon the jet
dynamics.  Expected observational implications of this model are
discussed in Section \ref{MI}.  Finally, concluding remarks are
contained in Section \ref{C}.

\begin{table}
\begin{center}
\begin{tabular}{lcc}
\hline
Helium Star Mass & $\MHe$ & $2\,\Ms$\\
Helium Star Radius & $\RHe$ & $2\times10^{11}\,\cm$\\
Orbital Separation & $a$ & $4\times10^{11}\,\cm$\\
Supernova Ejecta Density & $\dsn$ & $10^{-2}\,\g/\cm^3$\\
Supernova Ejecta Velocity & $\vsn$ & $10^3\,\km/\s$\\
\hline
Neutron Star Mass & $\MNS$ & $1.4\,\Ms$\\
Neutron Star Radius & $\RNS$ & $10^6\,\cm$\\
Black Hole Mass & $\MBH$ & $10\,\Ms$\\
\hline
Jet Lorentz Factor & $\Gamma$ & $20$\\
Jet Plasma $\sigma$ & $\sigma$ & $10^4$\\
\hline
\end{tabular}
\caption{A number of model parameters are listed together with some of
the jet characteristics obtained in the following sections.}
\label{quantities}
\end{center}
\end{table}

\section{Jet Formation} \label{JF}
A high power jet (or highly collimated relativistic outflow) is a
central feature of most $\gamma$-ray burst models.
Observational constraints require that the luminosity of a jet with
opening angle $\sim \Gamma^{-1}$ (where $\Gamma$ is the bulk Lorentz
factor) be on the order of
\begin{equation}
L_{\rm jet} \simeq \frac{1}{4\Gamma^2} L_{\rm iso}
\simeq 6\times10^{47} \left(\frac{\Gamma}{20}\right)^{-2}
\left(\frac{L_{\rm iso}}{10^{51}\,\erg/\s}\right)
 \,\erg/\s\,,
\label{JF:jet_luminosity}
\end{equation}
where $L_{\rm iso}$ is the isotropic equivalent luminosity of the
burst.  For a burst of duration $T$, this corresponds to a total
energy budget of
\begin{equation}
E_{\rm tot} \simeq 6\times10^{49} \left(\frac{\Gamma}{20}\right)^{-2}
\left(\frac{L_{\rm iso}}{10^{51}\,\erg/\s}\right)
\left(\frac{T}{100\,{\rm s}}\right)
\,\erg \,.
\label{JF:burst_energy}
\end{equation}

The observational features of the model presented here depend only
upon the existence of a suitably luminous and electromagnetically pure
jet (see, \eg, section \ref{PE:IfJT}) in the vicinity of the
supernova, and are otherwise independent of the mechanism by which
such a jet is produced.  Nonetheless, for concretness, here the jet is
presumed to be the result of accretion driven electromagnetic torques
upon a central compact object.  Therefore, possible mechanisms for the
production of such a jet are discussed below in the cases where the
compact object is a black hole or a neutron star.  However, this is in no means
meant to be an exhaustive discussion of the problem of jet formation,
which is beyond the scope of this paper.

\subsection{Black Hole Binaries}
A considerable literature exists regarding the production of
jets via the electromagnetic extraction of angular momentum from
accreting black holes
\citep[see, \eg,][and references therein]{Blan-Znaj:77,McKi-Gamm:04,Komi:04,Levi:05}.
The maximum energy that may be extracted by this method is determined by the
irreducible mass, $M_{\rm ir}$:
\begin{align}
E_{\rm tot}
&= M_{\rm BH} c^2 \left( 1 - \frac{M_{\rm ir}}{M_{\rm BH}} \right) \nonumber\\
&\simeq \frac{1}{8} j^2 M_{\rm BH} c^2\nonumber\\
&\simeq 10^{55} j^2\,\erg\,.
\end{align}
where $j$ is the dimensionless angular momentum of the black hole, and
the expansion is appropriate for $j\ll1$
\citep[see, \eg,][]{Misn-Thor-Whee:73}.  Comparing this to the energy
required to power a burst (\cf equation \ref{JF:burst_energy}) requires
$j\gtrsim0.01$, and thus puts a rather weak limit upon the spin of the
black hole.

In the model presented here, the torque upon the black hole is due to
a substantially super-Eddington accretion flow resulting from the
accretion of the supernova ejecta.  An estimate for the mass accretion
rate in this scenario is simply given by the Bondi-Hoyle rate,
\begin{align}
\dot{M}
&\simeq
\pi r_{\cal BH}^2 \dsn \vsn
\simeq
4 \pi \dsn \left(\frac{G \MBH}{c^2}\right)^2 \left(\frac{\vsn}{c}\right)^{-3} c \nonumber \\
&\simeq
2\times10^{29}\,\g/\s\,,
\label{JF:BH:mass_accretion_rate}
\end{align}
where the Bondi-Hoyle radius is given by
\begin{equation}
r_{\cal BH}
=
\frac{2 G \MBH}{\vsn^2 + c_s^2}\,,
\end{equation}
in which the sound speed ($c_s$) may typically be ignored compared to the
ejecta bulk velocity.  The resulting density in the vicinity of the
horizon is then
\begin{equation}
\rho
=
\frac{\dot{M}}{4 \pi r_h^2 c}
\simeq
\left(\frac{\vsn}{c}\right)^{-3} \dsn\,,
\end{equation}
where $r_h$ is the horizon radius.  This implies a magnetic field
strength on the order of
\begin{align}
\BBH &\simeq \sqrt{\dsn c^2} \left(\frac{\vsn}{c}\right)^{-3/2}\nonumber\\
&\simeq 2\times10^{13}\,\G\,,
\end{align}
presumably created via the magneto-rotational instability (MRI)
\citep{Hawl-Balb:95}.
The rotational energy may then be electromagnetically tapped by, \eg,
the Blandford-Znajek process \citep{Blan-Znaj:77}.  Recent numerical
simulations of accreting black holes have produced Poynting flux
dominated jets with typical luminosities of
\begin{align}
L_{\rm BH}
&\simeq \Omega_{\rm BH}^4 \BBH^2 \frac{1}{c^3}
\left(\frac{G\MBH}{c^2}\right)^6\nonumber\\
&\simeq 2\times10^{48} j^4\,\erg/\s\,,
\end{align}
where $\Omega_{\rm BH} = j c / 2 r_h$ is a measure of the angular
velocity of the black hole \citep{McKi:05}.  Therefore, rapidly
rotating black holes \cite[$j\gtrsim0.8$ for the black hole parameters
given here, \cf][]{Gamm-Shap-McKi:04} are easily capable of
producing a Poynting dominated jet of sufficient luminosity.

This scenario is similar to the central engine models for the
collapsar scheme \cite[see, \eg,][]{Poph-Woos-Frye:98}, the primary
difference being the accretion rate.  In both cases, the accretion
flow is neutrino cooled, and thus able to reach substantially
super-Eddington (\cf~with $10^{18}\,\g/\s$) accretion rates.  However,
whereas the typical collapsar accretion rates are on the order of
$1 M_\odot\,\s^{-1}$, here they are a considerably lower
$10^{-4} M_\odot\,\s^{-1}$.

\subsection{Neutron Star Binaries}
A second, and perhaps more speculative, mechanism by which the jet might be formed
involves a rapidly rotating neutron star.  As with the black hole, in
this scenario the energy is provided by the spin of the star.  The
total rotational energy available is
\begin{align}
E_{\rm tot} &= \frac{1}{2} I_{\rm NS} \Omega_{\rm NS}^2
\simeq \frac{1}{5} \MNS \RNS^2 \Omega_{\rm NS}^2
= \frac{G \MNS^2}{5 \RNS} \omega^2\\ \nonumber
&\simeq 10^{53} \omega^2 \,\erg \,,
\end{align}
where $\omega$ is the angular velocity in units of the breakup
velocity.  In order to be sufficient to power a burst this already
requires that $\omega \gtrsim 0.03$ and hence the neutron star must be
rapidly rotating ($P\lesssim 15\,\ms$) consistent with observations of
millisecond pulsars.

As with the black hole, the accretion of the supernova ejecta may
mediate the extraction of the rotational energy.  In the neutron star
case, the Bondi-Hoyle accretion rate is given approximately by
\begin{equation}
\dot{M} \simeq 5\times10^{27}\,\g/\s\,,
\end{equation}
at which, as in the black hole case, neutrino cooling dominates
\citep{Chev:89,Frye-Benz-Hera:96}.
Near the neutron star, this results in a density on the order of
\begin{equation}
\rho
\simeq
\frac{\dot{M}}{4\pi\RNS^2 v_r}
\simeq
10^4 \beta_r^{-1} \,\g\,\cm^{-3}\,,
\end{equation}
where $v_r \equiv \beta_r c$ is the infall velocity near the neutron
star surface.  This implies an accretion magnetic field strength on the
order of
\begin{equation}
\BNS \simeq 4\times10^{12} \beta_r^{-1/2}\,\G\,.
\end{equation}
Note that this is a lower limit as the radial infall velocity may be
substantially less than the speed of light.  This field will reconnect with
the native stellar field on time scales comparable to the infall time
scale \citep{Ghos-Lamb:79b}.  Therefore, it may not be necessary for the
native field of the neutron star to initially be dynamically
significant for it to subsequently strongly couple the star to the disk.

The resulting spin-down torque has been studied primarily in the
context of propellor accretion flows
\citep[see, \eg,][]{Ghos-Lamb:79b,Eksi-Hern-Nara:05,Roma-Usty-Kold-Love:04}.
Unfortunately, the magnitude of the spin down-torque is not a settled
issue.  Nonetheless, recent numerical and analytical efforts have
implied that
\begin{equation}
\mathcal{M} \simeq B_{\rm d}^2 R_{\rm d}^2 v_{\rm esc} \eta \left(1-\eta\right)\,,
\end{equation}
where $R_{\rm d}$ is the radius at which the magnetic field is
dominated by the inertia of the disk, $B_{\rm d}$ is the magnetic
field at this radius, $v_{\rm esc}$ is the escape velocity from this
radius, and $\eta$, the so-called fastness parameter, is the
rotation frequency of the neutron star in units of the Keplerian
velocity at the dissipation radius
\citep{Roma-Usty-Kold-Love:04,Eksi-Hern-Nara:05}.  Clearly the minimum
dissipation radius is the radius of the neutron star, and thus the
maximum {\em possible} luminosity is
\begin{equation}
L_{\rm NS max}
\simeq \BNS^2 \RNS^2 c\,\omega \left(1-\omega\right)\,,
\end{equation}
which is sufficient to drive a burst provided that
$B\gtrsim10^{13}\,\G$ and $\omega$ is near unity ($P\sim1\,\ms$).  The
former is satisfied if $\beta_r\simeq0.1$.  The latter must be
satisfied for a propellor to operate if the dissipation radius is near
the surface, placing a more stringent constraint upon the spin of the
star.  It should be noted that such a high spin may be expected in
neutron star--helium star binaries if the helium star's hydrogen
envelope was expelled due to accretion by the neutron star or common
envelope evolution, as is commonly thought to be the case in such
systems \citep{Beth-Brow:98}.

In the standard propeller theory the dissipation radius is assumed to
be approximately the Alfv\'en radius, where the pressure from the
neutron star's native field balances the accretion ram pressure.
However, in the hyperaccreting millisecond pulsar case
(considered here) the dominant magnetic field is the accretion flow
field.  Therefore, this case may be more analogous to accreting black
holes (in which the dissipation radius is roughly that of the horizon)
than to accreting pulsars.

The manner in which the dissipated energy is transported out of the
system is not clear a priori.  Numerical simulations of propellors
have found the development of a magnetic chimney, suggestive of the
development of a jet, suggesting that some fraction of the
energy leaves in a Poynting flux \citep{Roma-Usty-Kold-Love:04}.
This interpretation is supported by the obvious parallels with the
accreting black hole case, in which a significant fraction of the
spin-down energy is channeled into a Poynting jet.  However, a
definitive answer will depend upon the details of the accretion flow
and the magnetic coupling to the star, and will likely have to await
detailed numerical simulations.

\subsection{Jet Emergence} \label{JF:JE}
In order to have observational consequences it is necessary for the
jet to emerge from the growing supernova.  This is analogous to the
problem of jet emergence in the collapsar model
\citep[see, \eg,][]{MacF-Woos-Hege:01} with a
reduction of the external medium density by eight orders of magnitude.
The ram pressure of the jet,
\begin{align}
P_{\rm jet} &\simeq \frac{\Gamma^2 L_{\rm jet}}{3\pi r^2 c}
\simeq \frac{L_{\rm iso}}{12\pi r^2 c} \\ \nonumber
&\simeq 1\times10^{15}
\left(\frac{L_{\rm iso}}{10^{51}\,\erg/\s}\right)
\left(\frac{r}{10^{12}\,\cm}\right)^{-2}
\,\erg/\cm^3 \,,
\end{align}
can be compared to the typical pressures in the supernova ejecta,
\begin{equation}
P_{\rm sn} \simeq \frac{1}{6} \dsn \vsn^2
\simeq 2\times10^{13}\,\erg/\cm^3 \,,
\end{equation}
(the thermal photon pressure will be comparable at the orbital
separations considered here) and hence the jet will easily escape the
ejecta.

\section{Prompt Emission} \label{PE}
The presence of a jet alone is insufficient to produce a $\gamma$-ray
burst.  Also required is a mechanism by which the considerable kinetic
energy flux of the jet can be converted into the observed prompt
emission.  Many such mechanisms have been discussed in the
literature.  However, for highly beamed emission, a minimum
requirement is that the jet be optically thin along the jet axis.
This may be accomplished in a number of ways, including clumpy jets.
However, in the context of the model considered here, in which a large
number of seed photons are available entrained in the supernova
ejecta, this immediately suggests inverse-Compton scattering as the prompt
emission mechanism.  This has the considerable advantage over internal
shock scenarios of being capable of converting the kinetic energy of
the jet into the prompt emission at efficiencies approaching unity
\citep[see, \eg,][]{Lazz-Ghis-Celo:99}.

\subsection{Photon Collimation} \label{PE:PC}
Due to the high bulk Lorentz factor, the scattered seed photons will
naturally be collimated to within $\Gamma^{-1}$ of the jet axis
\citep[see, \eg,][]{Bege-Siko:87}.  However, in addition, there are
optical depth effects which will also serve to beam the scattered
photons.

In the rest frame of the jet electrons, the Thomson depth is given by
\begin{equation}
\tau = \int \sigma_T n_e c \d t' \,.
\end{equation}
When transformed into the lab frame this gives
\begin{align}
\tau &= \int \sigma_T n_e c \Gamma \left(1-\beta\cos\theta\right) \d t
\\ \nonumber
&=\int \sigma_T n_e \Gamma \left(1-\beta\cos\theta\right) \d r \,,
\end{align}
where in both cases $n_e$ is the proper electron number density.
For the two limiting cases of across the jet ($\theta\gg\Gamma^{-1}$)
and along the jet ($\theta\lesssim\Gamma^{-1}$), the optical depth is
\begin{align}
\tau_\parallel &= \frac{1}{2\Gamma}\sigma_T n_e r && \theta=0 \nonumber\\
\tau_\perp &= 2\sigma_T n_e r && \theta=\frac{\pi}{2}
\,,
\end{align}
where $n_e \propto r^{-2}$ (changing the power law index changes
$\tau_\parallel$ by factors of order unity) was assumed and a jet
width of $2 r/\Gamma$ was used.  Therefore, despite the considerable
difference in scale length, the optical depth along the jet is a
factor of $\Gamma$ less than that across the jet.  As a result, it is
possible to have a jet which is optically thin to photons within $\sim
\Gamma^{-1}$ of the jet axis while being optically thick to all
others.  This provides an additional collimating mechanism and
explains why it is possible to Compton scatter a large number of the
seed photons incident on the jet while allowing the scattered burst
photons to escape.

\subsection{Implications for Jet Type} \label{PE:IfJT}
The requirement that at interesting radii $\tau_\parallel<1$
limits the baryon loading of the jet.  The power associated with ions
in the jet is
\begin{align}
L_{\rm ions} &\simeq \frac{\pi r^2}{\Gamma} n_e m_p c^3
\simeq \frac{2 \pi r}{\sigma_T} m_p c^3 \tau_\parallel
\\ \nonumber
&\simeq 10^{-4} \left(\frac{\Gamma}{20}\right)^2
\left(\frac{r}{10^{11}\,{\rm cm}}\right)
\tau_\parallel L_{\rm jet}\,.
\end{align}
Hence, the condition that the jet be optically thin along it's axis at
$r\sim10^{11}\,{\rm cm}$ requires that in the jet,
\begin{align}
\sigma \equiv \frac{L_{\rm jet}}{L_{\rm ions}} - 1 \gtrsim 10^{4}\,,
\end{align}
where $\sigma$ is the ratio the Poynting flux to the kinetic energy
flux of the baryons.  Therefore, the jet must be extremely
electromagnetically dominated. It should be noted however that
observational precedent for such an electromagnetically dominated
outflow exists in the context of the Crab pulsar in which
$\sigma\sim10^6$ at these radii \citep[see, \eg,][]{Vlah:04,Komi-Lyub:04}.

The magnetic fields within the jet required to generate a Poynting
flux with the $\gamma$-ray burst luminosities will typically be of order
\begin{align}
B &\simeq \sqrt{2 \frac{L_{\rm iso}}{r^2 c}}\nonumber\\
&\simeq 3\times10^9
\left(\frac{L_{\rm iso}}{10^{51}\,\erg/\s}\right)^{1/2}
\left(\frac{r}{10^{11}\,\cm}\right)^{-1}\,\G \,,
\label{B_estimate}
\end{align}
for $r\gg \RNS$\footnote{The conical structure of
the jet will not necessarily extrapolate down to the acceleration
region which is expected to occur on a scale of many gravitational
radii.  If it were this would suggest magnetic field strengths on the
order of $10^{15}\,\G$, as is suggested in magnetar models of
$\gamma$-ray bursts as opposed to the far more conservative
${\rm few}\times10^{13}\,\G$ discussed here.}.
The resulting synchrotron cooling time for an electron with Lorentz
factor $\gamma$ in the frame of the jet is then
\begin{align}
t_{\rm syn} &\simeq \frac{\gamma m_e c^2}{P_{\rm syn}}
\simeq 6\pi\frac{m_e c^2}{\sigma_T c B^2} \gamma^{-1}
\\ \nonumber
&\simeq 10^{-10} 
\left(\frac{L_{\rm iso}}{10^{51}\,\erg/\s}\right)^{-1}
\left(\frac{r}{10^{11}\,{\rm cm}}\right)^2
\gamma^{-1}\,\s\,,
\end{align}
hence these electrons may always be treated as cold in this frame.

\subsection{Compton Seed Photons}
From momentum conservation, the energy of a photon after a single
Compton scatter is given by
\begin{equation}
\epsilon_f = \frac{1-\beta\cos\theta_i}
{1-\beta\cos\theta_f+
\epsilon_i\left(1-\cos\Theta\right)/\Gamma m_e c^2}
\epsilon_i\,,
\end{equation}
where $\theta_{i,f}$ are the initial and final photon propagation angle
with respect to the jet axis, and $\Theta$ is the angle between the
initial and final photon propagation direction.  For photons scattered
to within $\Gamma^{-1}$ (as expected from relativistic beaming and
optical depth collimation), the resulting energy is given
approximately by
\begin{equation}
\epsilon_f = \min\left(2 \Gamma^2 \epsilon_i,\Gamma m_e c^2\right) \,.
\end{equation}
Therefore, in order to generate a burst with an isotropic luminosity
of $L_{\rm iso}$, a luminosity of
$L_{\rm iso}/8\Gamma^4\simeq10^{45}(\Gamma/20)^{-4}\,\erg/\s$ of seed photons
are required to impinge upon the jet.

There are two sources of seed photons, the thermal photons from the
supernova itself and those produced by strong shocks driven into the
ejecta by the jet.  In the first case, due to the ejecta's high
Thomson depth, these will be entrained in, and in thermal equilibrium
with, the ejecta.  Thus, assuming the ejecta cools adiabatically as it
expands, the photon temperature will be
\begin{equation}
T_\gamma
\simeq \frac{m_p \vsn^2}{3 k} \left(\frac{r}{\RHe}\right)^{-2}
\simeq 4 \left(\frac{r}{\RHe}\right)^{-2}\,\keV\,,
\end{equation}
and hence the typical photon energy is on the order of $1\,\keV$ for
an orbital separation of $\gtrsim 2 \RHe$ which would be
expected if the Helium star is filling its Roche lobe.  The associated
energy density of photons is
\begin{equation}
u_T = a T_\gamma^4 \sim 2\times10^{16} \left(\frac{r}{\RHe}\right)^{-8}
\sim 10^{14}\,\erg/\cm^3\,,
\end{equation}
(where here $a$ is the Stefan-Boltzmann constant and not to be
confused with the orbital separation) and thus the luminosity of
thermal seed photons available to the jet is approximately
\begin{equation}
L_T \sim u_T \vsn \frac{2 r^2}{\Gamma}
\sim 10^{45}\left(\frac{r}{10^{12}\,\cm}\right)^2
\left(\frac{\Gamma}{20}\right)^{-1}\,\erg/\s \,,
\end{equation}
where clearly this depends upon the alignment of the jet relative to
the orbital plane through the dependence of $u_T$ upon the jet
position.  In the second case, the kinetic energy of the ejecta is
thermalised at the jet boundary providing a seed photon luminosity of
roughly
\begin{align}
L_S
&\sim \frac{1}{2} \dsn \vsn^3 \frac{2 r^2}{\Gamma}
\\ \nonumber
&\sim 10^{45} \left(\frac{r}{10^{12}\,\cm}\right)^2
\left(\frac{\Gamma}{20}\right)^{-1}\,\erg/\s \,,
\end{align}
which is comparable to $L_T$.  Depending upon the shock structure and
location these may also be thermalised.    Due to its weaker
dependence upon $r$ ($\dsn\propto r^{-3}$) this contribution will
dominate at large radii. The total luminosity of seed photons entering
the jet is then simply
\begin{equation}
L_{\rm seed} = L_T + L_S \,,
\end{equation}
which is sufficient.

\subsection{Implications for Jet Lorentz Factor} \label{PE:IfJLF}
Recent simulations suggest that in vacuum it would be expected that
the jet would accelerate due to magnetic stresses until it was no
longer electromagnetically dominated, \ie, $\Gamma$ would approach
$\sigma\sim10^4$ \citep{Spit-Aron:04}\footnote{Note that this is in contrast to
  many of the theoretical estimates in the literature in which
  behaviour expected by one-dimensional models is
  $\Gamma\rightarrow\sigma^{1/3}$ \citep[see, \eg,][]{Mich:69,Gold-Juli:70,Bege-Lee:94,Besk-Kuzn-Rafi:98}.
  \citet{Spit-Aron:04} claim that the discrepancy is largely due to
  the various artificial assumptions that these early efforts
  necessarily adopted, including spherical symmetry, split-monopole
  field configurations, and the region of applicability of the
  force-free approximation.}.  If the seed photons are indeed
from a thermal distribution with $T_\gamma \sim 1\,\keV$ as argued in
the previous sections, the up-scattered emission would peak at a few
$\GeV$, far higher than is observed. However, in the situation under
consideration here, the Compton scattering will induce a drag on the
jet, substantially reducing its velocity.  This will manifest itself
differently depending upon whether or not the jet is optically thin
along its axis.

When it is optically thin, the rate at which energy is lost by the jet
to the up-scattered photons at a given radius is simply given by the
rate at which seed photons enter the jet and their subsequent
up-scattered energy, \ie,
\begin{equation}
\frac{\d L}{\d r} = -2\Gamma^2 \frac{\d L_{\rm seed}}{\d r} \,,
\end{equation}
where $L_{\rm seed}$ as a function of radius is defined by
\begin{equation}
L_{\rm seed}(r) = \int_{r_0}^r \left( u_T + u_S \right) \vsn
\frac{2 r}{\Gamma} \d r \,,
\end{equation}
and $u_S$ is the energy density of seed photons produced in the shocks
(of order $\dsn \vsn^2/2$).  Since $\d L/\d r$ is negative definite it
will generally produce a deceleration in the jet.

In contrast, when $\tau_\parallel > 1$ the up-scattered photons cannot
escape.  They will consequently rescatter, introducing a second term
to the energy loss:
\begin{equation}
\frac{\d L}{\d r} = -2\Gamma^2 \frac{\d L_{\rm seed}}{\d r}
- 2 L_{\rm seed} \frac{\d \Gamma^2}{\d r} 
= -\frac{\d}{\d r} 2 \Gamma^2 L_{\rm seed} \,.
\end{equation}
Unlike the optically thin case, for strongly decelerating jets this
can produce a positive accelerating force.  This provides an effective
mass loading of the jet, and since $\sigma$ is so high, will determine
the maximum $\Gamma$ reached, \ie,
\begin{align}
\Gamma &\simeq \frac{L_{\rm jet}}
{2 \Gamma^2 L_{\rm seed}\vert_{\tau_\parallel=1}}
\simeq \frac{L_{\rm iso}}{8 \Gamma^3 \left(u_T + u_S\right) \vsn r^2}
\nonumber\\
&\longrightarrow \Gamma \simeq
30 
\left(\frac{L_{\rm iso}}{10^{51}\,\erg/\s}\right)^{1/4}
\left(\frac{r}{10^{11}\,{\rm cm}}\right)^{-1/2} \,.
\label{Lorentz_limit}
\end{align}
Hence if the jet becomes optically thin near
$r=2\times10^{11}\,\cm$, $\Gamma\simeq20$ as assumed thus far.
After $\tau_\parallel$ drops below unity the jet will continue to
decelerate until it exits the ejecta or stalls.

The Lorentz factors considered here are substantially lower than those
implied by the typical resolution to the ``compactness problem''
\citep[which are $\gtrsim {\rm few}\times10^2$, see, \eg,][]{Lith-Sari:01}.
The runaway pair production associated with the ``compactness problem'' is
not present here as a result of the low energy of the seed photons
themselves.  In the jet frame, the seed photons have
energy $\sim \Gamma \epsilon_{\rm sn}$, where $\epsilon_{\rm sn}$ is the
typical seed photon energy in the lab frame.  Since in the jet frame
the electrons are cold, and thus all of the photon scattering is elastic,
the maximum energy in any photon-photon collision is
$2\Gamma\epsilon_{\rm sn}$, which for $\Gamma\sim20$ and
$\epsilon_{\rm sn}\sim1\,\keV$ is insufficient to pair produce.

Nonetheless, for a thermal seed photon distribution a high energy tail
will exist, including photons with $\epsilon > m_e c^2/\Gamma$ and
hence will be able to pair produce.  The number density of such
photons accumulated in the jet frame by the end of the burst is
roughly given by
\begin{equation}
n_\epsilon \simeq g \Gamma \frac{u_T}{k T_\gamma} \frac{\vsn}{c}
\left(\frac{m_e c^2}{\Gamma k T_\gamma}\right)^2
\exp\left(-\frac{m_e c^2}{\Gamma k T_\gamma}\right)\,,
\end{equation}
where the factor of $\vsn/c$ accounts for the fact that in the jet the
photons are free streaming and $g\simeq0.15$ is a normalisation
factor.  The resulting optical depth to pair production is then
$\sigma_T n_\epsilon r/\Gamma$, where $r$ is the radius in the lab
frame \citep[see, \eg,][]{Lith-Sari:01}.  Therefore, the lepton
density at the end of the burst as a result of photon annihilation is
$n_{e^{\pm}} \simeq \sigma_T n_\epsilon^2 r/\Gamma$, with an associated
optical depth along the jet of
\begin{align}
\tau
&\simeq
\frac{1}{2\Gamma} \sigma_T n_{e^\pm} r \nonumber\\
&\simeq
\frac{1}{2} \left(\sigma_T n_\epsilon \frac{r}{\Gamma}\right)^2
\nonumber\\
&\simeq
3\times10^{-3}\,,
\end{align}
for $r\simeq10^{12}\,\cm$.  Note that since pair annihilation has been
ignored, this is only an upper limit.  Thus, it may be safely
concluded that pair production is insignificant during the the burst.

In general, both the jet and the seed photon density and temperature
will be expected to have radial structure.  In the jet this is due to
competition between the magnetic stresses and the Compton drag.  In
the seed photons this is due to the adiabatic cooling of the supernova
ejecta.  A direct result of this structure is that, despite beginning
with thermal seed photons, the time integrated spectra can have the
observed broken power law shape where the break energy could
indeed be interpreted as the temperature of the up-scattered seed photons
when $\tau_\parallel\simeq1$, \ie, roughly
$2\Gamma^2\,\keV\sim{\rm few}\times10^2\,\keV$.  For the case where
the seed photon temperature has a radial power-law dependence this has
already been explicitly shown to be the case by
\citet{Ghis-Lazz-Celo-Rees:00}.  However, in this scenario,
since the density of seed photons depends upon distance from the
helium star, different orientations of the jet with respect to the
orbital plane will lead to substantial changes in the spectral slopes
of the integrated emission.  Thus, because the compact object is
expected to have suffered a kick during its birth, the considerable
variation observed in burst spectral slopes would be expected.

\subsection{Population Statistics} \label{PE:PS}
For a given beaming angle
($\sim \Gamma^{-1}$) approximately $10^{-10} \Gamma^2$
bursts occur per galaxy per year \citep{Schm:01}.  This implies that the formation rate
(${\cal R}_{\rm CO-He}$) for the progenitor systems satisfy
\begin{equation}
{\cal R}_{\rm CO-He} \gtrsim 4\times10^{-8} \left(\frac{\Gamma}{20}\right)^2
\,{\rm galaxy^{-1} yr^{-1}} \,.
\end{equation}
There is a considerable literature which addresses the formation rates
of neutron star--neutron star binaries and neutron star--black hole
binaries.  Because in both cases it is believed that these are produced
by the evolution of compact object--Helium star binaries, the formation
rates of the former place lower limits upon the formation rate of the
latter.  Therefore, the progenitor formation rate must be compared to
${\cal R}_{\rm NS-NS} \simeq 10^{-6}$ to $5\times10^{-4}\,{\rm
  galaxy^{-1} yr^{-1}}$
\citep{Kalo-etal:04} and
${\cal R}_{\rm NS-BH} \gtrsim 10^{-4}\,{\rm galaxy^{-1} yr^{-1}}$
\citep{Beth-Brow:98}.
Hence, there are more than enough presumed compact object--Helium star
products to account for the number of $\gamma$-ray bursts observed.
It should also be noted that the formation rate of compact binaries is
expected to be considerably smaller than the formation rate of the
progenitor binaries due to the possibility of unbinding the
progenitor as a consequence of the ensuing supernova, and the fact
that the black hole--black hole binaries have been ignored.

\section{Afterglow} \label{Ag}
A consequence of electromagnetic domination in the jet is the absence
of internal shocks as long as the jet remains relativistic.  However,
as the jet cools, the flow will become increasingly hydrodynamic.
When $\Gamma$ is of order unity, strong internal shocks may be
expected to develop.  At this point the internal kinetic energy of
the jet can be thermalised, producing a moderately relativistic fireball,
which can then be analysed within the highly successful standard
fireball afterglow model (see, \eg, \citealt{Mesz:02};
\citealt{Koni-Gran:02} argue that many of the diverse afterglow
phenomena can be naturally produced if the afterglow is produced
inside of a pulsar-wind bubble).  Due to the
velocity gradient at the end of the jet, when enough matter
accumulates a Compton thick head will develop, shutting off further
prompt emission.  This can be expected to occur over the time scale for
the jet to proceed from the radius at which $\tau_\parallel\simeq1$
($\sim10^{11}\,\cm$) to the radius at which $\Gamma\sim1$
($\sim10^{12}\,\cm$), which for the scenario considered here is
approximately $30-100\,\s$. After this time, the jet will continue to
pump energy into the growing fireball at its head.  Only when the
rotational energy of the compact object is sufficiently exhausted, or
accretion ceases, will the jet cease and the fireball expand under its
own pressure.  For a neutron star this also can be expected to occur
over a comparable time scale as the prompt emission, while for a black
hole this can continue considerably longer (see section \ref{JF}).

\section{Model Implications} \label{MI}
This model has a number of direct observational implications discussed
below, many of which have already been detected.

\subsection{Burst Substructure}
The fact that the ejecta is Thomson thick and inhomogeneous leads to
considerable burst substructure.  In this case seed photons will not
be continuously available, but rather will enter the jet in bunches
over timescales ($\delta t$) associated with the inhomogeneity length
scales ($\ell$), \ie,
\begin{equation}
\delta t \simeq \frac{\ell}{\vsn} \sim 1\,\s\,,
\end{equation}
implying
\begin{equation}
\ell\sim10^3\,\km\sim10^{-3}\RHe\,.
\end{equation}
This will occur when an ejecta clump impacts the jet and is
subsequently either sheared apart, releasing the entrained photons, or
forms strong shocks and thus thermalising its bulk kinetic energy.

To lowest order, the rate of supplied seed photons may be treated as
uniform over $\delta t$ and vanishing otherwise.  In this case, for a
single clump, the number of seed photons within the jet will evolve
according to
\begin{equation}
\frac{\d N_{\rm seed}}{\d t} =
\left\{
\begin{aligned}
\frac{\cal N}{\delta t} - \frac{\tau_b}{\delta t} N_{\rm seed} && 0<t<\delta t\\
- \frac{\tau_b}{\delta t} N_{\rm seed} && \delta t < t\\
\end{aligned}
\right.
\,,
\end{equation}
where ${\cal N}$ is the total number of seed photons in the clump and
\begin{equation}
\tau_b = \tau_\perp \frac{\Gamma c \delta t}{2 r}\,,
\end{equation}
which is expected to be $\sim 0.3$ at the point where most of the
emission occurs.  The solution is trivially found to be
\begin{equation}
N_{\rm seed} =
{\cal N} \e^{\tau_b t/\delta t}
\left\{
\begin{aligned}
\frac{t}{\delta t} && 0<t<\delta t\\
1 && \delta t < t\\
\end{aligned}
\right.
\,.
\end{equation}
Since nearly every seed photon that enters the jet will be up-scattered
to $\gamma$-ray energies, this also provides the expected light curve
for a single subpulses
\begin{equation}
\frac{\d N_\gamma}{\d t} = f \frac{\tau_b}{\delta t} N_{\rm seed} \,.
\end{equation}
where $f$ is the fraction of singly scattered photons and is
typically of order unity.  This has the fast rise--exponential decay
structure observed \citep[\cf][]{Norr-etal:96}.  Furthermore, the
ratio of the exponential decay time to the (approximately) linear rise
time is $\tau_b^{-1} \sim 3$, as observed \citep{Norr-etal:96}.  As
the jet Lorentz factor increases these subpulses would be expected to
be systematically more symmetric for a given inhomogeneity scale in
the supernova ejecta (which would not be expected to vary
considerably).  Thus, more energetic bursts would be expected to
produce more symmetric subpulses, which has also been observed
\citep{Norr-etal:96}.

\subsection{Time Lags and Energy Dependent Subpulse Widths}
Photons at energies below $2\Gamma^2 T_\gamma$ will be due to both
the low energy tail of the seed photon distribution and multiple
scatters both within the jet and at the jet boundaries.  The
high bulk Lorentz factor of the jet implies that the scattering angles
will typically be on the order of $\Gamma^{-1}$.  Therefore,
for each encounter
\begin{equation}
\epsilon_f \simeq \frac{\epsilon_i}{1 + \Gamma \epsilon_i/m_e c^2} \,.
\end{equation}
Many encounters may be approximated by integrating the equation
\begin{equation}
\frac{\d\epsilon}{\d N_{\rm scat}} \simeq \frac{\epsilon}{1 + \Gamma \epsilon/m_e c^2} - \epsilon \,,
\end{equation}
to give
\begin{equation}
N_{\rm scat} \simeq \frac{m_e c^2}{\Gamma}
\left(\frac{1}{\epsilon}-\frac{1}{\epsilon_{\rm peak}}\right)
-\ln\left(\frac{\epsilon}{\epsilon_{\rm peak}}\right)
\sim \frac{m_e c^2}{\Gamma \epsilon}\,,
\end{equation}
and thus $\epsilon \sim N_{\rm scat}^{-1}$ in the limit of many scatters.  Due to the additional
path length traversed by these photons, they will lag behind the
single scattered photons by a time $\propto N_{\rm scat}$.  In addition, since
the scattered photons are performing a biased random walk, they will
spread in time $\propto N_{\rm scat}^{1/2}$.  This gives the following scalings
\begin{equation}
\Delta t_{\rm lag}
\propto \Gamma^{-1} \epsilon^{-1}\,,
\end{equation}
and
\begin{equation}
\delta t \propto \Gamma^{-1/2} \epsilon^{-1/2} \,,
\end{equation}
where the former is consistent with the observed anti-correlation between the
time lags and the overall luminosity of the burst \citep{Band:97},
and the latter is in rough agreement with the observed relation $\partial\ln\delta
t/\partial\ln\epsilon \simeq -0.4$ \citep{Feni-etal:95}.
The relevant time scales for the time lag can be estimated directly by
noting that the jet width is roughly
\begin{equation}
\frac{r}{\Gamma} \sim 5\times10^9
\left(\frac{\Gamma}{20}\right)^{-1}
\left(\frac{r}{10^{11}\,\cm}\right) \,\cm \,,
\end{equation}
and hence the lower energy emission will lag by
$N_{\rm scat} (r/\Gamma c)\sim0.2N_{\rm scat}\,\s$.  If a majority of the low energy
photons are produced deeper within the jet (where the Compton depth is
higher), this time scale can decrease by an order of magnitude.  In
either case this is again consistent with observations of burst pulse time
lags \citep{Band:97}.

\subsection{Spectral Softening}
As the jet evolves, its axial optical depth will increase and the
position at which the majority of the emission arises from will move
outward towards regions of lower Lorentz factor.  Near the end of the
burst the Lorentz factor at the $\gamma$-ray photosphere is expected
to be no more than a few.  As a result, the emission can be expected
to soften considerably as the burst proceeds.  The rate at which this
softening occurs depends upon the radial structure of the jet as
discussed in section \ref{PE:IfJLF} and would be expected to vary
considerably between bursts.

\subsection{Peak--Inferred Isotropic Energy Relation} \label{PIIER}
\citet{Amat-etal:02} \citep[and more recently][]{Amat:04} have
reported a correlation between the peak spectral energy and the
inferred isotropic energy of bursts, namely
$\epsilon_{\rm peak} \propto E_{\rm iso}^{1/2}$.
In the context of
the model presented here, this would be naturally expected if the
dominant factor in the variation between bursts was the maximum jet
Lorentz factor.  This is not unexpected if the helium star supernovae
are similar, implying that scatter in the orbital
separation and compact object parameters, both of which enter most
significantly into the determination of $\Gamma$, produces the
variability amongst bursts.  Then, in terms of the typical seed photon
energy ($\epsilon_{\rm sn}$), the peak $\gamma$-ray energy is
\begin{equation}
\epsilon_{\rm peak} \simeq 2 \Gamma^2 \epsilon_{\rm sn} \,.
\end{equation}
From equation (\ref{Lorentz_limit}) it is clear that
\begin{equation}
L_{\rm iso} \simeq 8 \Gamma^4 \left(u_T+u_S\right) \vsn r^2
\propto \epsilon_{\rm peak}^2\,,
\end{equation}
where the last proportionality holds if the typical ejecta densities
and velocities and the radius at which the jet becomes optically thick
are similar amongst bursts.  This may be trivially inverted to yield
the observed relation.

\subsection{X-ray Flash Characteristics}
For jet Lorentz factors $\sim\text{few}$, the peak of the emission
would occur in the X-rays, and thus this model provides a natural
explanation for X-ray flashes.
Note that this is neither a structured nor uniform jet model
\citep[\cf][]{Ross-Lazz-Rees:02,Lamb-Dona-Graz:04}.  In the former,
viewing an azimuthally structured jet from different angles provides
the peak--inferred isotropic energy relation.  In the latter, the
distribution is in the opening angle of the jet.  In both it is
assumed that there is little scatter in the total energy of the burst.
In contrast, as seen in equation (\ref{Lorentz_limit}) with the same
assumptions made in the previous subsection, in the scenario presented
here the energy released in the form of $\gamma$-rays is expected to
scale as
\begin{equation}
L_{\rm jet} \propto \Gamma^{-2} \,.
\end{equation}
However, in this as well as in the normal bursts, this energy is
expected to be subdominant relative to the energy in the supernova
itself.

Since X-ray flashes share many of the temporal and spectral features
of $\gamma$-ray bursts, including lying upon the peak--inferred
isotropic luminosity relation of \citet{Amat:04}, it is
tempting to interpret them as simply subluminous bursts.  Recently,
the bolometric energy of the X-ray flash \mbox{XRF 020903} has been measured
and was indeed found to be similar to typical $\gamma$-ray bursts
despite having a substantially lower inferred isotropic luminosity
\citep{Sode-etal:04}.

\subsection{Late Light Curve Variability}
Despite the large amount of energy released in the supernova, the
binary does not necessarily become unbound (indeed, it must not if
this is a viable formation scenario for double pulsars).  Therefore,
in systems with a sizable amount of ejecta remaining in the vicinity
of a nascent compact remnant, variability due to the occasional
accretion by the companion compact object may be expected after the
supernova becomes optically thin, and thus late in the light curve.
This may explain the late time ($\sim50\,{\rm days}$ after the burst)
variability observed in the residual light curve of \mbox{SN 2003dh},
which has a rough time scale of a few days, implying an orbital
separation of $\sim10^{12}\,\cm$, in agreement with what would be
expected in this model \citep{Math-etal:03}.

\subsection{Polarisation}
\citet{Lazz-Ross-Ghis-Rees:04} have shown that inverse-Compton
scattering by jets can produce large degrees of linear polarisation
($\sim100\%$ in some cases).
This is a direct result of the relativistic aberration of the seed
photons in the jet frame and is a function of the viewing angle of
the jet, nearing unity for $\theta\sim\Gamma^{-1}$.   Therefore,
jet models in which the prompt emission is due to inverse-Compton
scattering will generally exhibit wide variation in the degree of
linear polarisation, ranging from zero along the jet axis to unity
along the jet edge.  As a consequence, the degree of linear polarisation
may be used to measure the jet viewing angle, breaking the degeneracy
between azimuthal jet structure and low luminosity in attempts to
utilise $\gamma$-ray bursts as cosmological standard candles.  
Since, in the model presented here, the spectral evolution
of each subpulse to lower energy is due to multiple scatterings, the
polarisation fraction would be expected to decrease roughly
exponentially with the number of scatters and thus the polarisation
should be a strong function of energy as well.

Note that as long as the seed photon frequency in the jet frame is much
greater than the cyclotron frequency, as is expected to be the case
when the jet is optically thin axially (see equation
\ref{B_estimate}), the magnetic field will not significantly effect
the polarisation properties of the prompt emission.

Presently, the only measurement of the prompt emission polarisation
involved GRB 021206 and was found to be $80\%\pm20\%$ by
\citet{Cobu-Bogg:03} (though this claim continues to be
controversial).  However, due to the large uncertainty this may be
consistent with both synchrotron self-Compton models and the model
presented here.  If this is confirmed in future bursts, this
would provide strong evidence for Comptonised jet models.

\section{Conclusions} \label{C}
Rough estimates of the consequences of supernovae in compact
object--helium star binaries have been used to suggest that such an
event is a viable candidate for $\gamma$-ray bursts.  Such events are
assumed to occur in the standard double pulsar evolution scenario.
Within this context a number of population calculations have been
performed, implying that indeed these events are likely to be frequent
enough to explain the number of bursts seen.  This model is capable of
explaining a number of the burst characteristics, including the
subpulse light curves and energy dependence, late time spectral
softening, and the peak--inferred isotropic energy relation.  In
addition, X-ray flashes and $\gamma$-ray bursts are naturally combined
into a single unified theory, as recently suggested by observations.

If the binary remains bound, as is necessarily the case for the double
pulsar formation scenarios, late time variability may be expected.
Such variability has been observed in the light curve of
\mbox{GRB 030329}/\mbox{SN 2003dh}.  While no clear periodicity is
discernible, the typically time scales are consistent with an orbital
separation $\sim 10^{12}\,\cm$ as expected by this model.  Future
observations of late time afterglow light curves can provide evidence
for the existence of binaries in $\gamma$-ray burst progenitors.

The prediction of significant degrees of polarisation in Compton
dragged jet models in general, and this model in particular, is
noteworthy.  In the absence of strong structure within the jet, this
would necessarily lead to significant variation in the degree of
polarisation, ranging from zero to unity.  Predictions for significant
degrees of polarisation which depend upon viewing angle has
significant implications for cosmological observations.  Even in the
absence of a detailed understanding of the azimuthal jet structure,
polarisation provides a simple way in which to identify the viewing
angle, removing this degeneracy.  Unfortunately, due to its
considerable uncertainty, the only measurement of prompt emission
polarisation is as yet unable to differentiate between
synchrotron self-Compton models and Comptonised jet models.  Thus,
clearly future polarisation observations are required as well.

Many of the observational predictions of this model are generic to
Compton dragged jets.  These include the spectrum
\citep{Ghis-Lazz-Celo-Rees:00}, polarisation
\citep{Lazz-Ross-Ghis-Rees:04}, subpulse lags, and subpulse energy
dependence \citep[\cf][]{Dado-Dar-DeRu:02}.  To a lesser extent, the Amati
relation, and thus the X-ray flash characteristics, is also generally
expected as long as the primary difference among bursts is the jet
Lorentz factor.

This model is also very similar to the collapsar, with the obvious
difference that in this case the compact object is outside the helium
star (which need not be a Wolf-Rayet star here).  As such, the most
immediate feature of the collapsar model, the association with
supernova, is present here as well.  However, it is not possible to
simply subsume this model into the collapsar by placing the compact
object within the helium star; the primary difficulty being that the
solution to the compactness problem would no longer apply due to the
considerable supply of MeV photons in the central regions of the
supernova.  As a result the jet would remain optically thick until
radii two orders of magnitude larger than those considered here.
Considering the likely fact that the supernova ejecta are travelling
radially from the compact object at this point, the mechanism proposed
here for producing the burst substructure will be unlikely to work.

Unique to this model is the manner in which the successes of
Comptonised jet and collapsar models are unified.  Furthermore, it
provides a natural class of progenitors, in which supernovae are
already believed to occur as a result of independent observational
evidence.  Thus, the results presented here may be regarded as both a
proposal for an new $\gamma$-ray burst mechanism and an investigation
into the observational consequences of what appears to be an
inevitable process given the currently observed double neutron star
systems.

\section*{Acknowledgements}
I would like to thank a number of people with whom I've had useful
discussions, including Jon McKinney, Ramesh Narayan, Andrew McFadyen,
Roger Blandford, Max Lyutikov, Yasser Rathore, Gerry Brown, and Ralph
Wijers.  I would also like to thank the anonymous referee who's
suggestions substantially improved this manuscript.  I would
especially like to thank Martin Rees for a number of insightful
conversations and generously hosting me during the time that much of
this work was completed.  This research was supported by NASA-ATP
grant NAG5-12032 and NSF RTN grant AST-9900866.

\bibliographystyle{mn2e.bst}
\bibliography{grb.bib}

\bsp

\end{document}